\documentclass[twocolumn,preprintnumbers,amsmath,amssymb]{revtex4}

\usepackage{array}
\usepackage{graphicx}
\usepackage{tabularx}
\usepackage{longtable}
\usepackage{dcolumn} 
\usepackage{bm}

\begin{document}

\title{An oxide thermal rectifier
}

\author{W. Kobayashi$^{\dagger}$}
\affiliation{Waseda Institute for Advanced Study, Waseda University, 
Tokyo 169-8050, Japan}

\affiliation{PRESTO, Japan Science and Technology Agency, Saitama 332-0012, Japan}

\author{Y. Teraoka}
\affiliation{Department of Physics, Waseda University, 
Tokyo 169-8555, Japan}

\author{I. Terasaki}
\affiliation{Department of Applied Physics, Waseda University, 
Tokyo 169-8555, Japan}

\date{\today}

\begin{abstract}

We have experimentally demonstrated thermal rectification as bulk effect. 
According to a theoretical design of a thermal rectifier, we have prepared 
an oxide thermal rectifier made of two cobalt oxides 
with different thermal conductivities, 
and have made an experimental system to detect the thermal rectification. 
The rectifying coefficient of the device is found to be 1.43, 
which is in good agreement with the numerical calculation.   

\end{abstract}

\maketitle

Electric current and heat current are fundamental in the electron transport.
The electric current is precisely controlled, and plays a crucial role in various 
electronic devices. 
Though the heat current is the counterpart of the electric current, 
it is not well controlled. 
A thermal rectifier is a promising device in which heat flows in a forward direction 
while it can hardly flow in the opposite direction as an analogue of the diode. Owing to the controllability 
of the heat current, the thermal rectifier can be utilized for future practical application such as 
a thermal transistor \cite{li2}, thermal logic gates \cite{wang1}, and a thermal memory \cite{wang2}. 

%    ----------   Figure 1   ----------    %
\begin{figure}[t]
\begin{center}
\vspace*{0cm}
\includegraphics[width=7cm,clip]{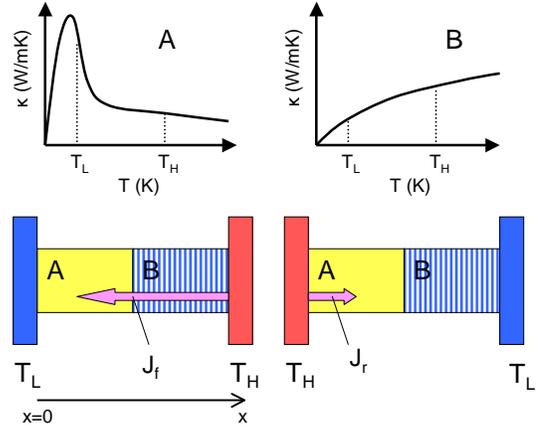}
\caption{(Color online) Top panels: Schematic figures of the thermal conductivity that is required for 
thermal rectification. Bottom panels: conceptual diagrams of the thermal rectification in bonded materials.
}
\end{center}
\end{figure}
%     -------------------------------     %

Recently, several theoretical calculations on the thermal rectification were reported 
\cite{terraneo, li1, peyrard, casati, yang, hu}. One-dimensional (1D) anharmonic-lattices model 
\cite{terraneo, li1}, and 1D-mass-graded-chain model \cite{yang} 
are examined for the calculation of thermal rectification in a bulk material 
or at an interface between two different materials. Surprisingly, a rectifying coefficient which 
is defined by the ratio of heat current in the forward direction to that in the opposite direction is 
predicted to be up to 2,000 \cite{li1}. 
In agreement with these theories, the thermal rectification is indeed observed in a carbon nanotube 
with mass gradient made by deposition of amorphous C$_6$H$_{16}$Pt \cite{chang1}, 
which can be advantageous to nano-scale devices. 
Peyrard has also proposed a simple design of thermal rectifier by juxtaposing two homogeneous materials 
with different temperature dependence of thermal conductivity \cite{peyrard}. 

In this letter, we experimentally demonstrate the thermal rectification in a device made of two perovskite cobalt 
oxides LaCoO$_3$ and La$_{0.7}$Sr$_{0.3}$CoO$_3$. 
The theoretical concept is shown in Fig. 1 \cite{peyrard}. 
We consider the heat current density $J$ through the bar made of materials A and B bonded at the center. 
$J$ is described by Fourier's law, 
\begin{equation}
J=-\kappa [x,T(x)]\frac{dT(x)}{dx},
\end{equation} 
where $\kappa $ represents the thermal conductivity as a function of position $x$ and temperature $T$. 
Here, we suppose that the material A exhibits a high $\kappa$ at low temperature $T_L$, and 
a relatively low $\kappa$ at high temperature $T_H$, while the material B displays the opposite property, 
as shown in the top panels of Fig. 1.  
Then, we provide two heat baths with $T_H$ and $T_L$. 
When the materials A and B are connected to the heat baths with $T_L$ and $T_H$, respectively, 
a total thermal conductance of the sample should be relatively high leading to 
a high heat current density $J_f$ through Eq. (1) (we call this direction the forward direction, 
see the lower left panel of Fig. 1). 
On the other hand, if the boundary conditions are reversed, 
the total thermal conductance becomes smaller than the other one, resulting in 
smaller heat current density $J_r$ than $J_f$ (the reverse direction, see the lower right panel of Fig. 1). 
Thus, the rectifying coefficient defined by the ratio of $\lvert J_f \rvert$ to $\lvert J_r \rvert$, 
\begin{equation}
R=\frac{\lvert J_f \rvert}{\lvert J_r \rvert}, 
\end{equation} 
should be above one. 
We have applied this concept proposed by Peyrard \cite{peyrard} 
to a sample made of LaCoO$_3$ and La$_{0.7}$Sr$_{0.3}$CoO$_3$ 
for the following advantages; (1) a small magnitude of the thermal conductivity ($\sim 5$ W/mK) compared with 
$\sim 100$ W/mK of conventional metals and alloys is efficient to maintain a large temperature gradient 
in the system, and (2) the difference of $\kappa (T)$ between the two oxides 
is found to be rather large \cite{berggold} so that a large rectifying coefficient is expected. 

%    ----------   Figure 2   ----------    %
\begin{figure}[t]
\begin{center}
\vspace*{0cm}
\includegraphics[width=5cm,clip]{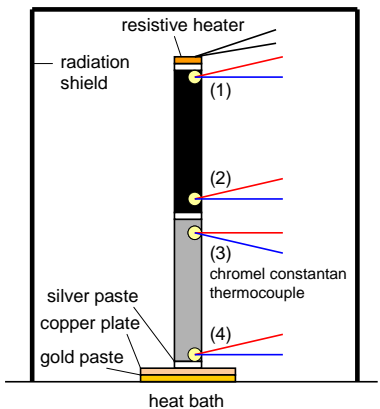}
\caption{(Color online) Schematic figure of the experimental configuration. The sample is 
connected to the heat bath with $T_L$ through a small thin plate of copper glued 
with silver paste. To monitor the temperature, four chromel-constantan thermocouples are 
glued at four specific points (1)-(4). 
}
\end{center}
\end{figure}
%     -------------------------------     %

Polycrystalline samples of LaCoO$_3$ (LCO) and La$_{0.7}$Sr$_{0.3}$CoO$_3$ (LSCO) 
were prepared by a solid state reaction. Stoichiometric amounts of 
La$_2$O$_3$, SrCO$_3$, and Co$_3$O$_4$ were mixed, 
and mixture was sintered at 1100 $^{\circ}$C for 12 h. 
Then, the product was finely ground, pressed into a pellet, and sintered at 1100 $^{\circ}$C for 24 h. 
X-ray diffraction of the sample was measured using a standard diffractometer with Fe K$\alpha$ 
radiation as an x-ray source in the $\theta -2\theta $ scan mode. Any impurity peaks were not 
detected. 

We made a measurement system as is schematically drawn in Fig. 2. 
Firstly, the two oxides of bar shape were bonded with silver paste with a high 
thermal conductivity (KYOCERA Chemical CT285) to reduce contact thermal resistance. 
The sample was annealed at 150 $^{\circ}$C for 0.5 h and 220 $^{\circ}$C for 2 h to dry the paste. 
The dimensions of the LCO and LSCO bars were $0.9\times 1.2\times 6.3$ mm$^3$ and 
$0.9\times 1.2\times 6.1$ mm$^3$, respectively, resulting in the total length of 12.4 mm. 
Then, a resistive heater (KYOWA strain gauges) with $\sim 130$ $\Omega$ including 
the resistance of electrical leads ($\sim 10$ $\Omega$) and a small piece of thin copper plate were 
attached at the top and bottom of the bonded sample using the silver paste, respectively. 
After the second annealing, the copper plate was glued on the 
cold head of a closed-cycle refrigerator with gold paste (Tokuriki Chemical Research). 
Then, four chromel-constantan thermocouples with 25-$\mu $m diameter were 
put at points (1)-(4) away from the ends of the LCO and LSCO bars by about 0.5 mm, 
and glued with varnish (GE7031, see Fig. 2). 
The sample was surrounded with a radiation shield made of copper plate to avoid the thermal radiation.

%    ----------   Figure 3   ----------    %
\begin{figure}[t]
\begin{center}
\vspace*{0cm}
\includegraphics[width=6.5cm,clip]{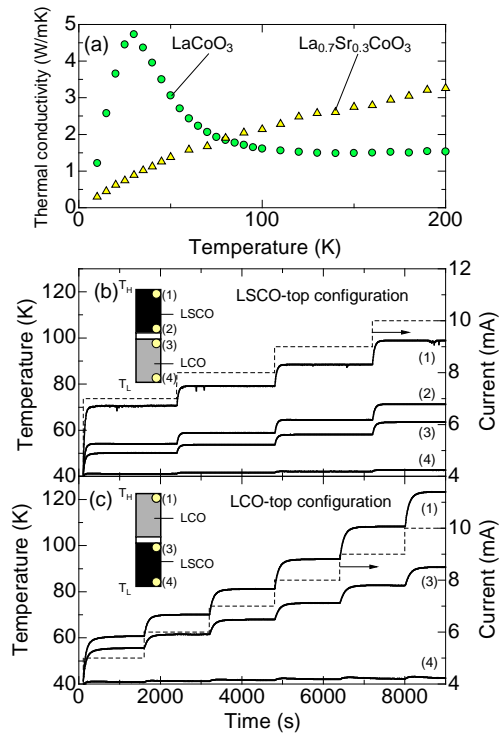}
\caption{(Color online) (a) The temperature dependence of the thermal conductivity 
of LaCoO$_3$ and La$_{0.7}$Sr$_{0.3}$CoO$_3$. 
Temperatures of the sample at the points (1)-(4) shown in the inset 
as a function of measuring time with 
(b) LSCO-top configuration, and (c) LCO-top configuration. 
}
\end{center}
\end{figure}
%     -------------------------------     %

The temperature of the sample was controlled from 3.5 to 200 K using a temperature controller 
(LakeShore 331S) with Cernox temperature sensor in the closed-cycle refrigerator 
evacuated down to 10$^{-4}$ Pa. 
For the thermal rectification measurement, the temperature of the cold head 
was kept at 40 K with a temperature stability of $\pm $10 mK. 
Then, the electric current was applied from 5 to 10 mA in steps of 1 mA into the resistive heater 
to make the heat current into the sample. 
We monitored the temperatures at the four points as a function of time and 
evaluated the thermal conductance of the sample. 
Radiation loss at 40 K with temperature difference of 60 K was evaluated to be below 1 \% 
using the Stefan-Boltzmann law. 
The leakage of the heat current from the four thermocouples and the electrical leads of the heater was 
also small ($\sim$ 1.5 \%). 
After the measurement for the forward direction, the heater and the copper plate 
were carefully removed, again glued at the opposite ends, and then the measurement 
was repeated for the reverse direction. 
The thermal conductivity was measured using a differential thermocouple made of 
chromel and constantan to detect a small temperature gradient of 1-2 K/cm.  

%    ----------   Figure 4   ----------    %
\begin{figure}[t]
\begin{center}
\vspace*{0cm}
\includegraphics[width=8cm,clip]{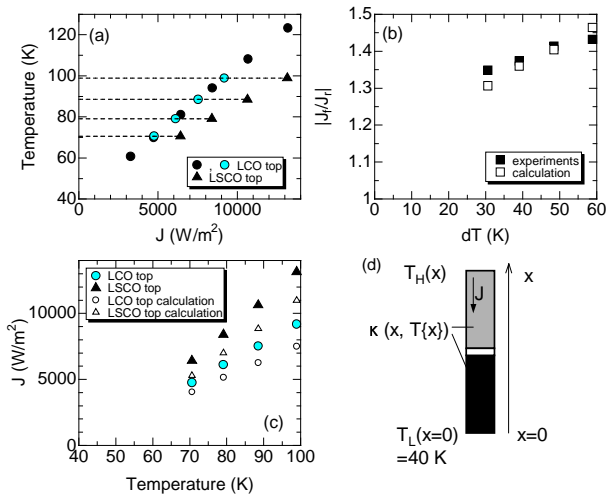}
\caption{(Color online) (a) The monitored temperature at the point (1) as a function of 
the heat current density $J$, 
(b) the measured and calculated rectifying coefficients $R$, 
(c) the measured and calculated heat current density $J$ as a function of the temperature at the point (1), and 
(d) schematic figure of the sample for the numerical calculation. 
}
\end{center}
\end{figure}
%     -------------------------------     %

Figure 3 (a) shows the temperature dependence of the thermal conductivity $\kappa $ 
of LaCoO$_3$ and La$_{0.7}$Sr$_{0.3}$CoO$_3$. 
Two samples exhibit different temperature dependences; 
$\kappa$ of LaCoO$_3$ is 1.5 W/mK at 200 K, and increases with decreasing temperature, 
while $\kappa$ of La$_{0.7}$Sr$_{0.3}$CoO$_3$ 
is 3.3 W/mK at 200 K and decreases with decreasing temperature. 
These data are similar to those reported by Berggold {\it et al} \cite{berggold}. 
Due to the grain boundary in our polycrystalline samples, the magnitude is somewhat smaller than 
their results for the single crystal samples. 

Figures 3(b) and (c) show the monitored temperatures at the points (1)-(4) of the forward and 
reverse directions as a function of time, respectively. As the forward direction, we chose LSCO-top configuration 
as shown in the inset of Fig. 3(b). 
We applied 7-10 mA and 5-10 mA into the resistive heater in steps of 1 mA 
for LSCO- and LCO-top configurations, respectively (see dotted lines of Figs. 3(b) and (c)). 
After reaching a steady state, the temperature was stable within $\pm $20 mK at each temperature. 
Note that the temperature at the point (4) 
increases only slightly, and that the contact thermal resistance between 
the sample and the heat bath at $T_L=40$ K is negligible. 
On the other hand, the temperature difference $\bigtriangleup T$ between points (2) and (3) was 
7.8 K in the LSCO-top configuration, which is rather large compared with 4.8 K 
evaluated from the thermal conductivity shown in Fig. 3(a). 
This shows that $\bigtriangleup T$ is about 3 K at the interface. Since $\bigtriangleup T$ between both ends of the 
sample is 58.9 K, the contact thermal resistance at the interface should be about 5 \% of 
the total thermal resistance of the sample. 
The most remarkable feature is that the monitored temperature at the point (1) in the LSCO-top configuration 
is substantially different from the one in the LCO-top configuration, 
though the applied current to the heater is the same. 
This clearly indicates different thermal resistances between the LSCO-top 
and LCO-top configurations.

Figure 4 (a) shows the monitored temperatures of the two configurations 
at the point (1) as a function of the heat current density $J$. 
The temperature is different between the two configurations 
in the presence of the same $J$. 
To make the same temperature at the point (1), $J$ was 
adjusted by controlling the electric current (see broken line and open circles). 
Obviously, $J$ is different between the forward and reverse directions 
in the presence of the same temperature difference $\bigtriangleup T$$ =$ $T_H -T_L$ showing the thermal rectification. 
As shown in Fig. 4(b), the rectification coefficient $R$ is evaluated to be 1.43 for 
$\bigtriangleup T$ $=$ 58.9 K. To our knowledge, this is the record at present. 
We would like to add a note that direction-dependent interfacial thermal resistance 
at the LCO/LSCO interface can be negligibly small. 
According to Refs. \cite{li1,hu}, different phonon bands between two materials cause the 
interfacial thermal rectification. Since the crystal structures of LaCoO$_3$ and 
La$_{0.7}$Sr$_{0.3}$CoO$_3$ are almost isostructural one another, 
these materials can have similar phonon bands leading to negligibly small interfacial thermal rectification. 

Lastly, we perform the numerical calculation of the heat current density $J$ 
and the rectifying coefficient $R$ using Fourier's law. 
From Eq. (1), the temperature distribution is given by 
\begin{equation}
T(x)=T_L+\int_{0}^{x}\frac{J}{\kappa [\xi ,T(\xi )]}{\rm d}\xi. 
\end{equation} 
To calculate $J$, the boundary condition $T$($x=13.4$ mm)=$T_H$= 
70.6, 79.1, 88.5, 98.9 K, and $T$($x=0$ mm)=$T_L=40$ K were applied, 
and the thermal conductivity data shown in Fig. 3(a) was used as $\kappa$[$\xi$, $T(\xi )$]. 
The calculated heat current density $J_{\rm calc}$ is shown in Fig. 4(c). 
$J_{\rm calc}$ well describes the experimental data. 
Using $J_{\rm calc}$, $R_{\rm calc}$ is obtained as shown in Fig. 4(b). 
$R_{\rm calc}$ also well reproduces the experimental value, 
which strongly indicates that the differences in $\kappa [x,T(x)]$ distribution 
are responsible for the observed thermal rectification. 
As proposed in Ref. \cite{peyrard}, the thermal rectification can also be achieved 
by nonlinear-lattice model and geometrical effect on a sample shape. 
Moreover, the thermal rectification can be realized in a narrow temperature range 
when one uses materials in a vicinity of a phase transition. 
The selection of the optimal material and the combination of the models will yield 
higher rectification coefficient. 

In summary, we have prepared a device made of cobalt oxides 
LaCoO$_3$ and La$_{0.7}$Sr$_{0.3}$CoO$_3$ according to 
a theoretical proposal of thermal rectifier, and have made an 
experimental system to detect the thermal rectification. 
The observed heat current density $J$ in the forward direction is obviously 
larger than that in the reverse direction leading to a large rectifying coefficient $R$ of 1.43 
with the temperature difference of 58.9 K between heat baths. 
We have calculated $J$ and $R$ using 
thermal conductivity data and found that the calculation is in quite good agreement with 
the experimental data. Therefore, we conclude that 
the observed thermal rectification is caused by a combination of two different materials 
whose thermal conductivities show different temperature dependences.  

We appreciate Y. Horiuchi, S. Shibasaki, and M. Abdel-Jawad for 
fruitful discussion and valuable comments. 
This study was partly supported by the Murata Science Foundation.

\end{document}